\documentclass[10pt,twocolumn,letterpaper]{article}

\usepackage[english]{babel}
\usepackage{multirow}
\usepackage{diagbox}
\usepackage{subcaption}

\usepackage{times}
\usepackage{epsfig}
\usepackage{graphicx}
\usepackage{amsmath}
\usepackage{amssymb}
\usepackage{enumitem}

\PassOptionsToPackage{hyphens}{url}
\usepackage[colorinlistoftodos]{todonotes}

\usepackage{comment}
\usepackage{verbatim,epsfig,times}
\usepackage{color}
\usepackage[export]{adjustbox}         
\usepackage[T1]{fontenc}

\usepackage[breaklinks=true,bookmarks=false]{hyperref}

\usepackage[ top = 1.50cm, bottom = 2.50cm, left = 2.00cm, right = 1.50cm]{geometry}
\date{\vspace{-5ex}}
\begin{document}

\title{Down for Failure: Active Power Status Monitoring}

\author{
 \begin{tabular}{ p{2.55cm} p{2.55cm} p{2.55cm} p{2.55cm} p{2.55cm} p{2.55cm} }
 \multicolumn{2}{ c }{Niloofar Bayat}&\multicolumn{2}{ c }{Kunal Mahajan}&\multicolumn{2}{ c }{Sam Denton} \\ 
 \multicolumn{2}{ c }{\normalsize{\textit{Department of Computer Science}}}&\multicolumn{2}{ c }{\normalsize{\textit{Department of Computer Science}}}&\multicolumn{2}{ c }{\normalsize{\textit{Department of Computer Science}}} \\ 
 \multicolumn{2}{ c }{\normalsize{\textit{Columbia University}}}&\multicolumn{2}{ c }{\normalsize{\textit{Columbia University}}}&\multicolumn{2}{ c }{\normalsize{\textit{Columbia University}}} \\ 
 \multicolumn{2}{ c }{\normalsize{niloofar.bayat@columbia.edu}}&\multicolumn{2}{ c }{\normalsize{sam.denton@columbia.edu}}&\multicolumn{2}{ c }{\normalsize{mkunal@cs.columbia.edu}} \\
 & & & & & \\
 & \multicolumn{2}{ c }{Vishal Misra}&\multicolumn{2}{ c }{Dan Rubenstein} & \\ 
 & \multicolumn{2}{ c }{\normalsize{\textit{Department of Computer Science}}}&\multicolumn{2}{ c }{\normalsize{\textit{Department of Computer Science}}}& \\ 
 & \multicolumn{2}{ c }{\normalsize{\textit{Columbia University}}}&\multicolumn{2}{ c }{\normalsize{\textit{Columbia University}}}& \\ 
 & \multicolumn{2}{ c }{\normalsize{vishal.misra@columbia.edu}}&\multicolumn{2}{ c }{\normalsize{danielr@columbia.edu}} &
 \end{tabular}
}


\maketitle

\begin{abstract}
Despite society's strong dependence on electricity, power outages remain prevalent. Standard methods for directly measuring power availability are complex, often inaccurate, and are prone to attack. This paper explores an alternative approach to identifying power outages through intelligent monitoring of IP address availability. In finding these outages, we explore the trade-off between the accuracy of detection and false alarms. 

We begin by experimentally demonstrating that static, residential Internet connections serve as good indicators of power, as they are mostly active unless power fails and rarely have battery backups. We construct metrics that dynamically score the reliability of each residential IP, where a higher score indicates a higher correlation between that IP's availability and its regional power. We monitor specifically selected subsets of residential IPs and evaluate the accuracy with which they can indicate current county power status.

Using data gathered during the power outages caused by Hurricane Florence, we demonstrate that we can track power outages at different granularities, state and county, in both sparse and dense regions. By comparing our detection with the reports gathered from power utility companies, we achieve an average detection accuracy of $90\%$, where we also show some of our false alarms and missed outage events could be due to imperfect ground truth data. Therefore, our method can be used as a complementary technique of power outage detection. 
\end{abstract}



\section{Introduction}\label{Introduction}
Since 1974, the total electricity consumption of the world increased at an annual rate of 3.4\%~\cite{ieastatistics}, reaching 20,200 TWh by 2015, of which 3758 TWh was consumed in the United States \cite{ieastatistics, eia}. Given the crucial role of electricity in society, power outages from cyber-attacks~\cite{case2016analysis} and natural disasters~\cite{ferc2012arizona} 
have devastating effects with economic, social, physical and psychological impacts. Blackouts in the United States due to weather-related outages have increased 5-10 times since the 1990s \cite{castillo2014risk}. Reports by the U.S. Department of Energy (DOE), EPRI, and LBNL have estimated \$30-\$400 billion per year in economic losses due to power outages \cite{castillo2014risk}.
The ability to detect such outages today relies on power monitoring and diagnostic systems that are typically realized through wired communications. However, due to the high cost of installing and maintaining highly reliable communication cables that are resilient to outages, they are not widely implemented today ~\cite{gungor2010opportunities}. 

These outages and concerns about existing detection infrastructure put America's aging, sprawling power grid system under the spotlight. In response, DARPA launched the Rapid Attack Detection, Isolation and Characterization Systems (RADICS) program, one of whose goals is maintaining situational awareness by providing accurate and timely information about the power grid's state before and after cyber-attacks~\cite{darparadics}.

In this paper, {\em we explore the efficacy of identifying regional (county-level) outages from IP probes}. We begin by identifying properties of IP addresses in each geographic region who are stable with the power, where we can infer power availability from their up/down status. To accurately monitor power status, these IPs need to be scanned frequently (on the order of a few minutes). Using IP probes has the added benefit that the location of the monitoring point can be far from the location where outages are being probed, and furthermore can be decentralized to provide additional resilience. Our approach works with both IPv4 and IPv6 address spaces, although here our work focuses on using IPv4 probes.

While IP probing has been used to detect outages, to the best of our knowledge, existing Internet-based probing methods~\cite{schulman2011pingin,quan2013trinocular,kopp2013controlled,cardona2013weather,dischinger2007characterizing,wong2007octant, Dainotti2014, Dainotti2012} focus on {\em general} Internet outages, regardless of their cause, power or otherwise. 
In contrast, we distinguish between outages that are not due to power failure and those which are, and hence perform a more in-depth analysis of the cause of probes that indicate some general notion of failure. A further contrast with prior work can be found in \S\ref{RelatedWork}. Additionally, our method performs its assessment on-line, with the ability to provide notifications of power outages as they are occurring. 

We demonstrate that static, residential IPs are most susceptible to power by comparing our probe results to data gathered from utility company reports~\cite{poweroutages} in over 1500 counties in the U.S. 
We present our process for selecting subsets of residential IPs to be our {\em watchlist}, which is dynamically updated over time such that the list includes those IPs that statistically are the most correlated with power availability. 
Our results show a correlation between our detected outages with outages reported by power utilities, whose data we use as a point of comparison. We show that during massive power outages such as the ones caused by Hurricane Florence, our method matches power company reports $90\%$ of the time, with false positive rate and false negative rates averaging below $10\%$. Furthermore, we highlight those instances where IP scans identified outages before power company data did, either because of human error or because power reports are delivered at coarser timescales.
 
 We summarize our contributions as follows:
\begin{itemize}
\item We design a process that identifies IP addresses who are stable and their up/down status is correlated with regional power.
\item We show how we design our scanning process to be sensitive to concerns about ICMP flood rate that scanning processes of this type produce. 
\item We demonstrate that we can distinguish power outages from general Internet outages whose causes may result from failures aside from power loss.

\end{itemize}

The rest of the paper is organized as follows. In \S\ref{DesignChallenges}, we describe the challenges we had to overcome in our design. \S\ref{Methodology} explains our pipeline in detail and describes our method to dynamically determine informative IPs. \S\ref{Evaluation} presents our preliminary results. In \S\ref{RelatedWork} we discuss related work. \S\ref{Discussion} will conclude the paper with a discussion of the implications of this work and describe our future work.

\section{Design Challenges} \label{DesignChallenges}

Using IP pings effectively to identify power outages requires us to address 3 sub-problems:

\begin{enumerate}
 \item How do we ensure that the IPs we monitor are good indicators of regional power availability? In short, we wish to scan IPs that are up when power is available and down when not.
 
 \item How can we differentiate between a lack of responsiveness from IPs due to power failure from other situations that might induce lack of responsiveness (e.g., BGP misconfigurations, DDoS attacks, Internet service provider outages, etc.)?
 
 \item How do we ensure that our scanning rates are robust enough to provide useful information concerning power outages without generating an overabundance of ICMP floods and background radiation?

\end{enumerate}

Before delving into our solutions to these problems in \S\ref{Methodology}, we describe these challenges in greater detail.

\subsubsection*{Which IP addresses to monitor} While only roughly 3.6\% of all $2^{32}$ possible IP addresses respond to pings~\cite{heidemann2008census}, of this subset, many IPs are not reliable indicators of power availability. In particular, we seek to avoid IP addresses that

\begin{itemize}
 \item are dynamically assigned: there are over 102 million dynamically IP addresses \cite{xie2007dynamic}, and while some may be good indicators of power, there are periods where such an address may be unassigned (and hence appear down) or assigned to a mobile device that can remain operational while regional power is down.
 
 \item belong to data centers and big companies that have power backups and will maintain connectivity during regional outages. Thus, their corresponding IP addresses might also remain operational during outages and hence are generally considered risky indicators.
 
 \item after ruling out the above two criteria, may historically exhibit unresponsiveness even when power is up. This can be for any variety of reasons. Generally, since such IP addresses show a lack of reliability during times that are not outages, we prefer to avoid using these IP addresses as indicators.
 
\end{itemize}
\begin{figure*}[ht!]
 \centering
 \centering
 \includegraphics[width=\linewidth]{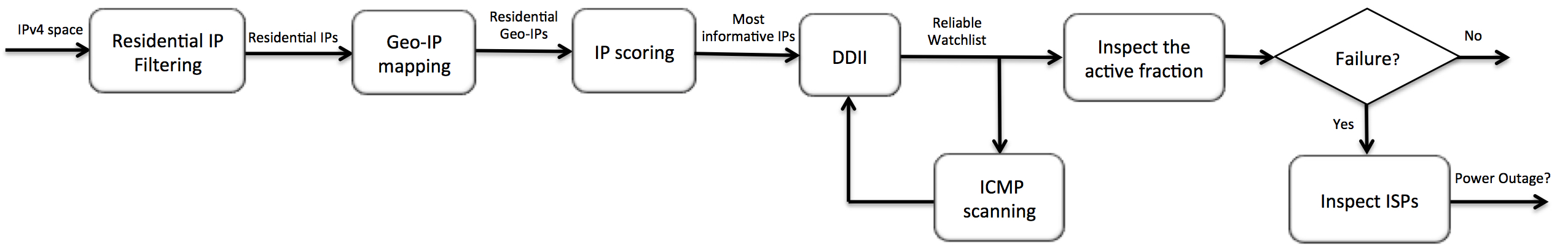}
 \caption{Our pipeline process}
 \label{fig:pipeline}
\end{figure*}

Our rationale is to generally avoid IPs in classes whose availabilities are unlikely to correlate with regional power status, build sampling histories of the remaining IP addresses with which we can construct a measure of reliability/con\-fi\-dence of each IP address, and weight our assessment of regional power using probes to those IPs that we judge to be more reliable indicators of power outage. Further details of this process are discussed in \S\ref{Methodology}.

\subsubsection*{Minimizing Internet background radiation} IPv4 and IPv6 address space both contain address space pollution, or unused blocks, which are mainly the result of environmental factors (e.g., misconfiguration, location), rather than algorithmic factors \cite{ wustrow2010internet}. These blocks will not be informative in our method, hence we need to avoid monitoring them to avoid causing unwanted traffic and to minimize Internet background radiation \cite{pang2004characteristics, wustrow2010internet}. While we historically assess the informativeness of each single IP address, the unused IP blocks will be ruled out in our automated system. Furthermore, since we stochastically select IPs with high reliability to probe, if the unused blocks start to be assigned to actual entities our automated system will cover them over time.

\subsubsection*{Impacting Internet service} Increasing the overall probe rate of IP addresses will, for the most part, help improve detection accuracy, but can also cause unintended anomalous behavior, in addition to generally being frowned upon. The same can be said for frequent probing of any specific IP address. Hence, we must consider how to effectively scan a region's collection of IP addresses that are good indicators of regional power without overloading the region as a whole, or overloading any particular subset of IP addresses~\cite{DoSA}. 

\subsubsection*{Bandwidth limitations} 
In addition to the aforementioned constraints on the probing rate, the source points of probes may impose their internal limits of probing capacity (we have such a restriction at our institution). We also must accommodate such source point limits as well, and if this is a bottleneck, determine how best to partition the available probing bandwidth among the set of regions being monitored. 

\section{Methodology and Design}\label{Methodology}
%
%
Figure \ref{fig:pipeline} depicts a pipeline process describing our overall approach that performs periodic IP scans to determine regional power availability. As mentioned in \S\ref{Introduction}, we focus on residential IP addresses for our power failure detection system as they are more susceptible to power failures and serve as a good indicator. We first explain an initializing stage in which we identify residential IP addresses (\S\ref{sec:resIP}) and map these addresses to geographical regions (\S\ref{sub:Geolocational}). We further refine this list of IPs into what we call a {\em residential watchlist} (\S\ref{sec:residential}) that is continually monitored at a slow rate. The residential watchlist is then further refined into a much smaller, dynamically adjusted {\em reliable watchlist} that contains IPs that are judged to be the best indicators of power availability. The process of forming and updating the reliable watchlist as well as the collection and interpretation of probe results from the watchlist is described in \S\ref{sub:DDII}.


\subsection{Residential IP determination}
\label{sec:resIP}

Residential IP addresses are the most susceptible Internet connections to power outages \cite{schulman2011pingin}. Henceforth, we study their behavior to detect power outages. There are several characteristics of IP addresses indicating whether or not they are residential. These characteristics mainly include the ISP names, port numbers, and security methods \cite{shodanreport}. We use Shodan \cite{matherly2009shodan} due to its capability for gathering a list of host IP addresses categorized by their ISPs, were we obtain IP blocks belong to the most popular residential ISPs \cite{dischinger2007characterizing}. Furthermore, to obtain only residential IP addresses, port numbers 110, 123, 161, and 5060 are avoided as they are often used for businesses or data centers 
\cite{shodanreport}. In addition to port numbers, security methods HTTP, FTP, ssh, telnet encompass non-residential IP addresses and should be avoided during queries. 
 Therefore, after obtaining IP blocks based on their ISP, we filter out port numbers and security methods well-known for non-residential connections. From this method, we obtain approximately 200,000,000 US residential IP addresses, which form the basis of our residential watchlist. Since the residential watchlist consists of CIDR blocks of IP addresses, there are some unassigned IPs hidden in our residential watchlist who do not respond to ICMP pings and we remove them in our next step of reliable watchlist formation.

\subsection{Geolocational mapping of IP addresses}\label{sub:Geolocational}
 After determining the residential IP addresses, we map them to their corresponding geographical location, resulting in a list of ``GeoIPs". Hence, we can gather information from each specific region by monitoring the IP addresses assigned there. We use the MaxMind City database for geo-IP coordinate information \cite{maxmind2006geoip}. MaxMind is up to 86\% accurate in mapping IP addresses within a region of 50 kilometer radius according to their support center \hyperlink{https://support.maxmind.com/geoip-faq/geoip2-and-geoip-legacy-databases/how-accurate-are-your-geoip2-and-geoip-legacy-databases/}{website}. We then utilize an API from \emph{Federal Communication Comission} (FCC) to determine the county of each IP address based on their coordinates \cite{FCCdata}. 

\subsection{The Residential Watchlist} 
\label{sec:residential} 

The residential watchlist is formed starting with the IPs identified above and is then further thinned to exclude any blacklisted IP addresses, which either belong to reserved IP addresses or are obtained through an opt-out mechanism where users submit IP addresses or blocks to be excluded.


Once thinned, the IPs in the remaining residential watchlist are scanned every 6 hours via ICMP pings by ZMap \cite{durumeric2013zmap} to assess their general availability over time. We keep updating the blacklist as we receive more requests from IP blocks to be excluded through our opt-out process, and if we find out about new IP blocks that potentially belong to residential IP addresses, we add them to our residential watchlist. 





 \subsection{Maintaining and Using the Reliable Watchlist: DDII method}\label{sub:DDII}

The residential watchlist provides a set of IPs we consider as potentially valid. However, not every IP on this watchlist is an equally informative estimate of power availability. We further refine the residential watchlist to a much smaller, dynamically adjusted {\em reliable watchlist} using a heuristic, DDII which \emph{dynamically determines informative IP addresses}. DDII performs five essential tasks:
 \begin{enumerate}
  \item It scores each IP on the residential watchlist based on the collected probe histories.
  
  \item At a slow period (every 6 hours), it probes the entire residential watchlist, building a historical recording that tracks if each IP on the residential watchlist responds to a probe.
  
  \item At a faster period (a varying time on the order of few minutes) it builds a much smaller {\em reliable watchlist} and additionally probes this smaller watchlist.
  
  \item If an unexpected fraction of reliable watchlist is down in each region, it detects a {\em failure} event in that region.
  
  \item In regions with failure, it assesses whether the failure corresponds to a power outage by inspecting IP responses within each ISP.
  
  \item It further adjusts scores of IPs on the residential watchlist and probing periods of regions based on the results of outage status.
  
 \end{enumerate}
 
 \subsubsection{IP Address Scoring}
 
 Each time an IP address is probed, we record the time of the probe and the probe result, the latter of which is a Boolean indicator of whether there was a response from that address. The score of the IP address is a simple exponential weighted moving average (EWMA) of these responses, defined recursively as 
 \begin{equation}\label{eq:EWMA}
 S_j(i) = S_j\left(i-1\right) (1-\alpha) + \alpha \sigma_j\left(i\right)
 \end{equation}
 where $S_j(i)$ is the score after $i$ probes of IP address $j$, and $\sigma_j(i)$ is the indicator of the $i$th probe of $j$, where we have $\sigma_j(i) = 1$ if IP address $j$ responds to our ping in the $i$th probe, and $\sigma_j(i) = 0$ otherwise. Note that the initial condition $S_j(0)$ has minimal impact on the score in the long-term, as a sample's bias fades at a rate of $(1-\alpha)^k$ after $k$ subsequent samples. In our current implementation, $S_j(0) =0.5$ and $\alpha=0.01$. These values initially populate an inconclusive score, but as we build a sufficient body of samples, then recent histories as a whole are given significantly more weight than older histories or just the most recent few samples.
 
 \subsubsection{Slow-Period Full Residential Watchlist Scans}
 
 Our experiments begin with no information about the availability of IP addresses on the Residential Watchlist. To seed IP's initial scores, we run scans every 6 hours over the entire residential watchlist. For these low-rate scans, we make a simplifying default assumption that power is always available, and that a non-response is likely due to an indication other than a power failure. While this assumption may induce a small error in assessing a node's availability independent of power failure, it generally suffices for its intention, which is to provide information about which IPs will most often be available when there is power.

 \subsubsection{Reliable Watchlist Scans}
 
 In addition to our low-rate scans above, we provide higher rate scans that cover each geographical region of interest. Each region has its rate at which these higher scans occur (discussed below in \S\ref{sub:failuremode}). Each time that region is to be scanned, a small (relative to the residential watchlist in that region) subset of IPs are selected as members of the reliable watchlist. This subset is scanned, and the results of that scan are used to assess the power status of the covered region (see \S\ref{sub:powerdetermine},\S\ref{sub:ISP}). If the hypothesis is that the region has power, then this round of scans is used to update the scores of the IPs on the reliable watchlist who were probed. Otherwise, the results of the scans are not incorporated into the IP's score $S_j$. Here, we intentionally exclude these scans from the score during perceived outage periods to limit the bias in the score of non-responses that occur due to power outages.
 
 \subsubsection{Reliable Watchlist Sizing}
 \label{sec:reliableIPsizing}
 
\begin{figure}[t]
 \centering
 \centering
 \includegraphics[width = \linewidth]{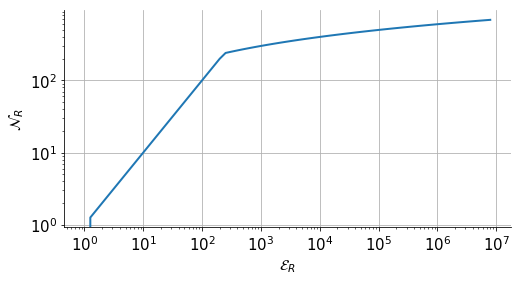}
 \caption{The relationship between size of reliable watchlist ($\mathcal{N}_R$) and the expected response probe rate in region $R$ ($\mathcal{E}_R$). }
 \label{fig:N_E}
\end{figure}

 For each region $R$, we define
 \begin{equation}
 \label{equ:Eval}
  \mathcal{E}_R = \lim_{i \rightarrow \infty} \sum_{j \in R} S_j(i),
 \end{equation}
 i.e., $\mathcal{E}_R$ is the expected rate of response to probes in region $R$ per scan. Note that the limit of $i\rightarrow \infty$ is the ideal number of scans to make sure the score of IP address $j$ has converged to its final value, but any large value of $i$ could be used for this purpose, have we used scan data for over a year to determine this score. A very small value of $\mathcal{E}_R$ decreases our confidence in our ability to infer power outage status in that region, as the number of sample points responding with availability is simply too small to make any statistically accurate claims. Hence, we do not track power status of regions $R$ for which $\mathcal{E}_R < 10$. Otherwise, the number of samples within an iteration of scanning of region $R$ is chosen to be
 \begin{equation}
 \mathcal N_R= 
 min\{\left \lfloor{\mathcal{E_R}}\right \rfloor, \left \lfloor{100 log_{10}(\mathcal{E_R}) }\right \rfloor\}.
 \label{eq:choice}
 \end{equation}
 Note that the requirement $\mathcal{E}_R \geq 10$ is an arbitrary option and we chose it as a result of a trade off between accuracy of detection and number of covered regions. 
 Figure \ref{fig:N_E} depicts the relationship between $\mathcal N_R$ and $\mathcal{E}_R$ in logarithmic scale. $\mathcal N_R$ grows linearly with respect to $\mathcal{E}_R$ for small values, and then after reaching a threshold, grows sub-exponentially. This is to provide sufficient diversity in the set of IPs being scanned, but not grow excessively large as to cause unnecessary traffic sent to a particular region. 
 
 \subsubsection{Reliable Watchlist Formation}
 
 After determining the number $\mathcal N_R$ of IPs to add to region $R$'s reliable watchlist, the reliable watchlist is reconstructed each time it is to be sampled. For a given sampling, the specific set of IPs comprising that sample's watchlist are chosen using a biased random selection process over the set of all IPs in the residential watchlist for that region. Specifically, each IP $j$ is selected for inclusion in the $i+1$st reliable watchlist with a probability proportional to its current score, $S_j(i)$. Hence, selection is biased toward IPs that have histories of being active when power is available. However, the randomness ensures that other IPs are also sometimes included within the reliable watchlist. This has two benefits: first, it reduces the overall sampling burden imposed on high-scoring IPs, and second, it provides more frequent assessment of lower scoring IPs than simply the 6-hour scans, as both frequent and our baseline 6-hour scans are used to update IP scores in case no failure is detected.

 \subsubsection{Regional Failure Determination}
 \label{sub:powerdetermine}
 
For a given scan of the reliable watchlist, we compare the actual result to the expected result, given the current makeup of the watchlist. Specifically, the current result is 
\begin{equation}
U_{\mathbb{R}}(i) = \frac{\sum_{j \in \mathbb{R}} \sigma_j(i)}{|\{j: j \in \mathbb{R}\}|},
\end{equation}
i.e., the average up/down status of all members on the reliable watchlist. The expected result is 
\begin{equation}
E_{\mathbb{R}}(i) = \frac{\sum_{j \in \mathbb{R}} S_j(i-1)}{|\{j: j \in \mathbb{R}\}|}.
\end{equation}
Where $\mathbb{R}$ is the reliable watchlist in region $R$. We compare the difference ($E_{\mathbb{R}}(i) - U_{\mathbb{R}}(i)$) and when this threshold exceeds a value $\tau$, we indicate a failure. $\tau$ is a parameter whose value depends on the scale of the failure we wish to determine. For instance, $\tau$ is small (e.g., $.01$ if we wish to register sub-regional failures), and can be larger (e.g., $0.5$) to register only large-scale outage events. It is possible to have multiple values of $\tau$ applied in parallel so that in parallel, we can detect small to large outages. In our study, we use $\tau=0.07$ as an indicator as to whether the current sample should be incorporated into the IP's scores. However, we vary our decision to report an outage based on varying values of $\tau$, depending on the size of outages that we are attempting to detect. Details regarding the selection of $\tau$ for different analyses of results are presented in the \S\ref{Evaluation}.

\subsubsection{Distinguishing Power Outages from General Internet Outages}
\label{sub:ISP}

A novel contribution of our work is further discerning power outages from more general Internet outages. Here, our analysis uses a heuristic that when a power outage occurs within a region, clients served by the various ISPs throughout the region should be similarly affected. In contrast, when the problem has an alternate cause, this typically affects only a subset of regional ISPs. Hence, we indicate a power outage only when all ISPs in the region are similarly negatively impacted by the apparent outage. In other words, upon failure detection in a region where ($E_{\mathbb{R}}(i) - U_{\mathbb{R}}(i)$) exceeds the value $\tau$, we inspect if the same condition holds within each ISP, in which case we report a power outage. 

Note that in some rare cases, outages across multiple ISPs could be caused by reasons other than power outages. For example, a shared submarine cable cut or nation wide Internet blackouts. 
However, since these events are generally very large, information about them would become available very quickly and even though our model would capture a general failure in these rare cases, after detection and taking precaution, we would know quickly that there had not been a grid failure. Furthermore, due of lack of data, distinguishing between power and Internet failures in these cases can be addressed as part of future work.

 \subsubsection{Reliable Watchlist Scan Frequency}
 \label{sub:failuremode}
 
 The rate at which a region's reliable watchlist is scanned depends upon the current hypothesis of power status within the region. Associated with each region is a counter $C_R$ that decrements every two minutes. When the counter reaches 0, $R$'s reliable watchlist is formed, a scan ensues, and the counter is reset. Let $V_i$ be the value to which it was reset during the $i$th scan. Then $V_{i+1}$ depends on the current detected status. In instances where no failure is detected in the region, $V_{i+1} = \min(5,V_i + 1)$, and if a failure was detected, $V_{i+1} = \max(1, V_i-1)$. Hence, we increase the rate of scanning gradually (up to a maximum of every $2$ minutes) during periods where an outage is suspected (to gather more accurate measurements during this anomalous period) and gradually decrease it (down to a minimum of every $10$ minutes) during periods where power is assumed available. 
 
 \subsection{Preliminary Measurements}
 
\begin{figure}[t!]
 \centering
 \centering
 \includegraphics[width=\linewidth]{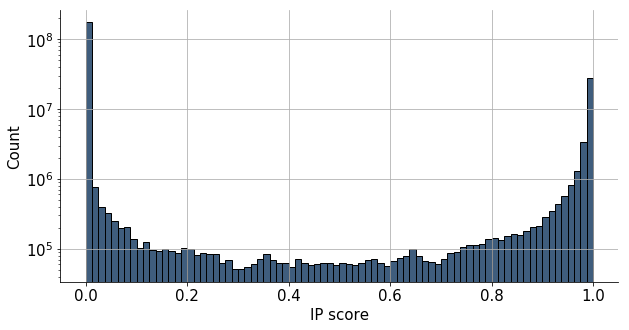}
 \caption{The distribution of IP scores among the residential IP addresses in the U.S. }
 \label{fig:IP_score}
\end{figure}

 Figure \ref{fig:IP_score} depicts the IP score distribution within our entire residential watchlist. The low IP scores represent IP addresses that do not tend to respond to our ICMP ping, whereas the IP addresses with high scores frequently respond to our pings. We observe that there are plenty of IP addresses within the residential watchlist that have very low scores and monitoring them would not provide us with much information about the power status. Hence, we devise our method so that we mainly focus on monitoring IP addresses with high scores. 
 
\begin{figure}[t!]
 \centering
 \begin{subfigure}[t]{\linewidth}
 \centering
 \includegraphics[width=\linewidth]{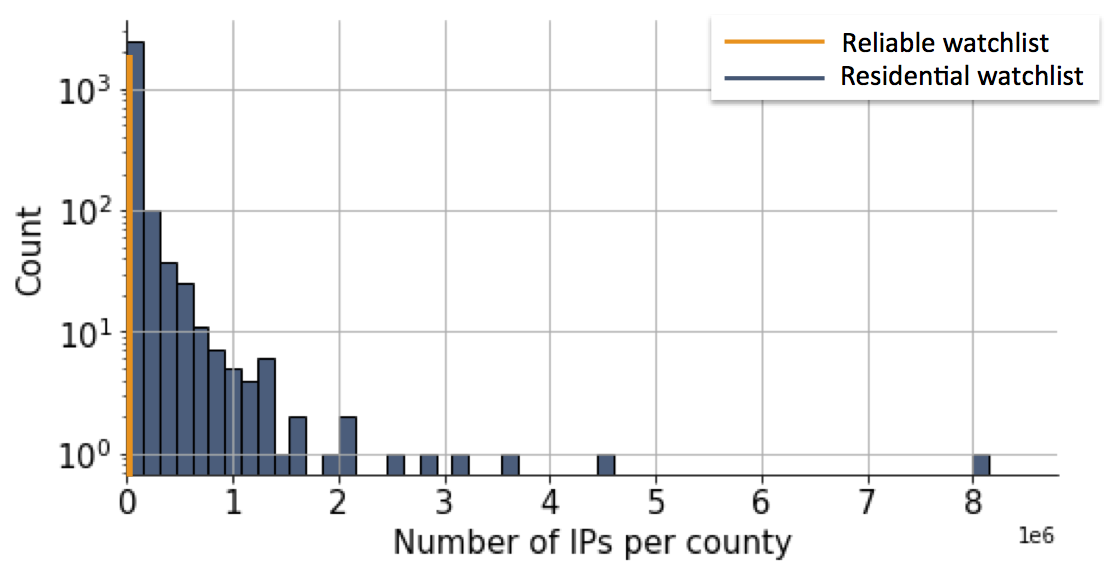}
 \caption{Residential watchlist}
 \end{subfigure} 
 \begin{subfigure}[t]{\linewidth}
 \centering
	\includegraphics[width=\linewidth]{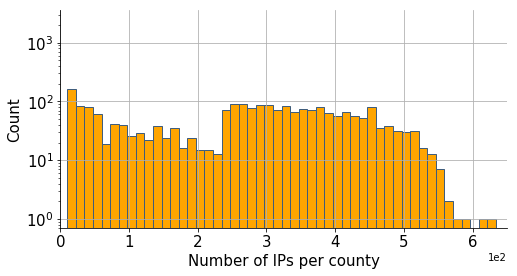}
 \caption{Reliable watchlist}
 \end{subfigure}
 \caption{distribution of number of IP addresses per county in (a) the residential watchlist and (b) the reliable watchlist.}
 \label{fig:residential2reliable}
\end{figure}

 Figure \ref{fig:residential2reliable} depicts the distribution of the number of IP addresses per county in (a) the residential watchlist and (b) the reliable watchlist. Note the difference in scale of the $x$-axis between these two figures (\ref{fig:residential2reliable} (a) the x-axis has scale $1e6$ while in \ref{fig:residential2reliable} (b) the scale is $1e2$). The number of IP addresses monitored by the reliable watchlists is generally four orders of magnitude smaller than their corresponding residential watchlists, reducing the ICMP ping flood rate, as well as our bandwidth utilization. Recalling that we do not monitor regions for which $\mathcal N_R < 10$, we do not include the respective county in our study. Therefore, there is a gap at $x=0$ in Figure \ref{fig:residential2reliable} (b). The reliable watchlist as a whole is also depicted in Figure \ref{fig:residential2reliable} (a) in orange color to permit a direct comparison between residential and reliable watchlist sizes.

\section{Evaluation}\label{Evaluation}
In this section, we evaluate our method using the utility reports of power outages within the U.S. \cite{poweroutages}, and Internet scan results generated by DDII with the average scanning period of less than $10$ minutes. Hence, any power outage that lasts over 10 minutes should be detected. We use the data from the utility reports as verification of our IP-probe based detection method over 1500 counties in the U.S.; counties in which we have both valid utility data and enough reliable IP addresses to monitor. 

\subsection{Existing Baseline: Utility Company Reports}
 The utility reports used from \hyperlink{poweroutages.us}{poweroutages.us} combine information from over 600 Utilities into one repository, making it the most complete source of power outage information currently available. Since this data is the best publicly available information and is not based on Internet scans, we use their data to validate the conclusions drawn by our process. We note that utility report data does itself has limitations as described in the data collection process \cite{utilitylimitations}: specifically, it does not include information of all utility companies, and its reliance on customer reports can introduce a human error that biases its conclusions. Going forward, our process could provide further validation and insight into these reports.

\begin{figure}[t]
 \centering
 \centering
 \includegraphics[width = \linewidth]{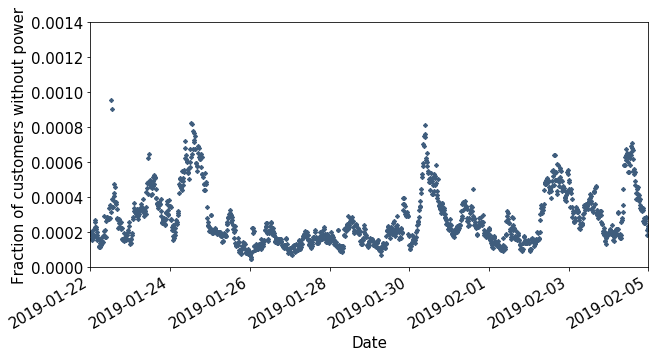}
 \caption{The utility reports of power loss within the U.S. from Jan 22, 2019 to Feb 5, 2019.}
 \label{fig:utility}
\end{figure}

 Figure \ref{fig:utility} depicts the fraction of outages in a period of two weeks (from Jan 22, 2019 to Feb 5, 2019) extracted from \hyperlink{poweroutages.us}{poweroutages.us}. Each region's outage reports are normalized with respect to their local time zone. There were no major outages in this period, and the fraction of outages reported follows a periodic diurnal pattern that peaks near mid-day. We hypothesize three reasons as to why this diurnal pattern might occur with drops at night:
\begin{itemize}
 \item Customers are less likely to report outages at night. 
 \item Fewer utility line workers are available to record and fixing power outages during the night.
 \item The smaller utilities that do not have smart meters may not record outages.
\end{itemize}


Although the power utility data we are using as ground truth might be inaccurate at times, especially for small events as many small utility companies do not report outage information online \cite{utilitylimitations}, it is useful as a sanity check for our method, as power utility data capture most minor and almost all major events. Note that the list of counties included in this dataset is static, the only thing that changes is the percentage of active customers.

\subsection{Implementation}
Our method is implemented on a 64-bit GNU/Linux machine with Intel Xeon 8C E5-2620 v4 2.1GHz processors and 128GB memory.
To establish the reliability of DDII, 
we use the heuristic proposed in Section \ref{Methodology}, where we randomly select among the IP addresses proportional to their scores, and we increase the rate of monitoring of counties during periods of suspected outages and simultaneously stop updating scores until the outage dissipates. 

\begin{figure}[t]
 \centering
 \includegraphics[width=\linewidth]{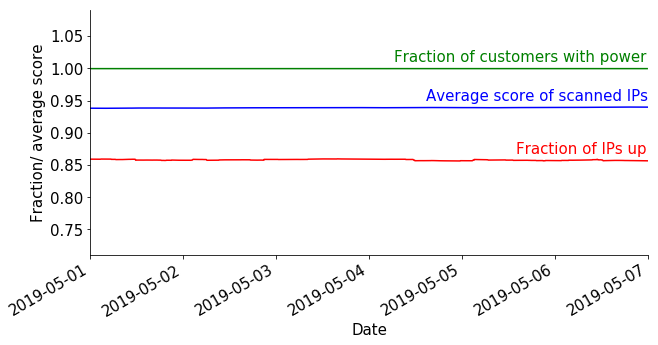}
 \caption{The average score and the fraction of active IP addresses, as well as the fraction of customers with power over a week. A moving average with 24-hour length is applied to the results for better visualization.}
 \label{fig:avg_score_all}
\end{figure}

Figure \ref{fig:avg_score_all} depicts a national view of three metrics of interest to our study, measured over a week-long period in which no major outages occurred. The top curve presents the fraction of customers receiving power as reported by the power company. Without any major outages, this value remains close to 1.0 (i.e., almost 100\% of customers have power). The next line is our average score of scanned IPs, depicting $E_{\mathbb{R}}(i)$ over time (see \S\ref{sub:powerdetermine}), and beneath that is the fraction of IPs currently reported as up, depicting $U_{\mathbb{R}}(t)$ over time (again see \S\ref{sub:powerdetermine}).

From Figure \ref{fig:avg_score_all} we observe that the fraction of active IP addresses on average is around $0.07$ lower than the average score of scanned IPs obtained from EWMA. Note that this unexpected bias of $0.07$ is due to a difference that EWMA achieves when compared with the traditional mean~\cite{tseng2003study}. While we currently address this bias by incorporating it into our assessment, future work will explore making appropriate adjustments that can remove such bias from the EWMA.

\begin{figure*}[t]
 \centering
 \begin{subfigure}[t]{0.3\linewidth}
 \includegraphics[width=\linewidth]{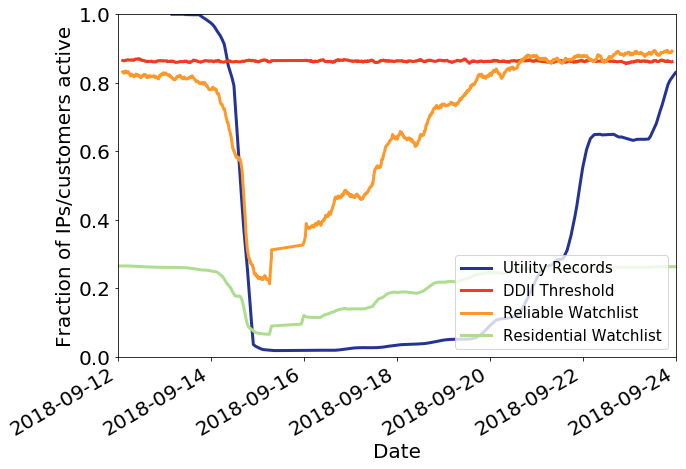}
 \caption{The active fractions in Robeson, North Carolina}
 \end{subfigure} 
 \begin{subfigure}[t]{0.02\linewidth}
 \includegraphics[width=\linewidth]{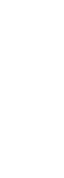}
 \end{subfigure} 
 \begin{subfigure}[t]{0.31\linewidth}
 \centering
	\includegraphics[width=\linewidth]{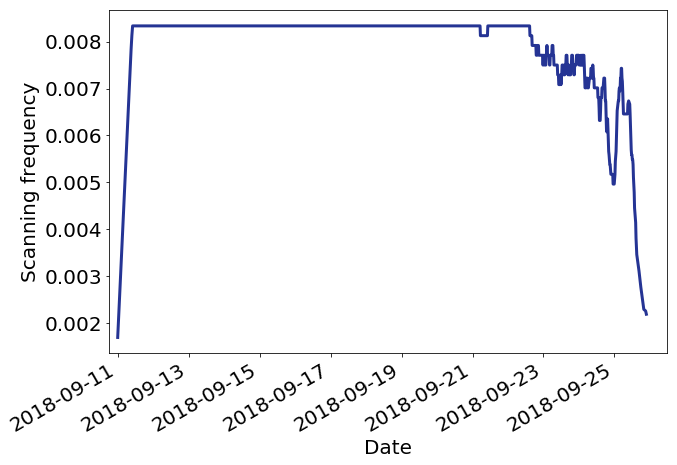}
 \caption{The scanning frequency in Robeson, North Carolina\\}
 \end{subfigure}
 \begin{subfigure}[t]{0.02\linewidth}
 \includegraphics[width=\linewidth]{Figures/filler.png}
 \end{subfigure} 
 \begin{subfigure}[t]{0.3\linewidth}
	\includegraphics[width=\linewidth, right]{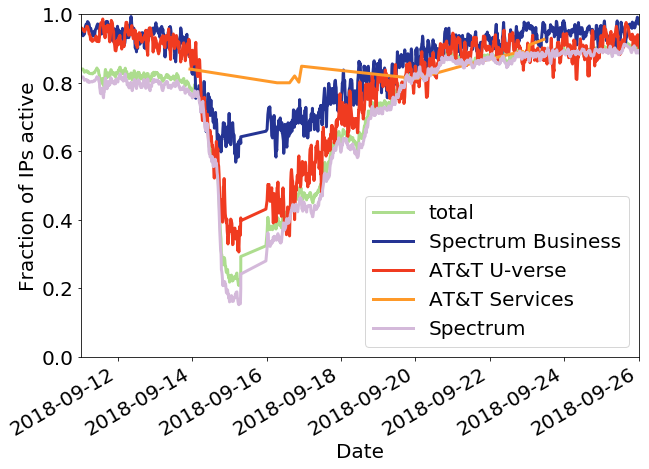}
 \caption{The active fractions for different ISPs in Robeson, North Carolina\\}
 \end{subfigure}
 \begin{subfigure}[t]{0.3\linewidth}
 \centering
	\includegraphics[width=\linewidth]{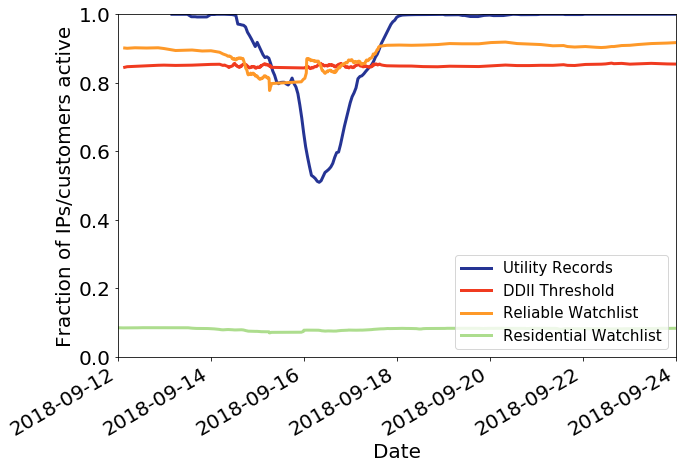}
 \caption{The active fractions in Lancaster, South Carolina \\}
 \end{subfigure}
 \begin{subfigure}[t]{0.02\linewidth}
 \includegraphics[width=\linewidth]{Figures/filler.png}
 \end{subfigure} 
 \begin{subfigure}[t]{0.31\linewidth}
 \centering
	\includegraphics[width=\linewidth]{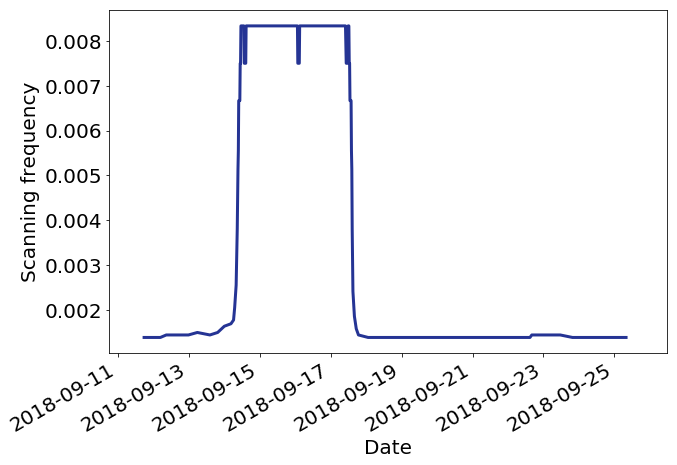}
 \caption{The scanning frequency in Lancaster, South Carolina}
 \end{subfigure}
 \begin{subfigure}[t]{0.02\linewidth}
 \includegraphics[width=\linewidth]{Figures/filler.png}
 \end{subfigure} 
 \begin{subfigure}[t]{0.3\linewidth}
 \centering
	\includegraphics[width=\linewidth]{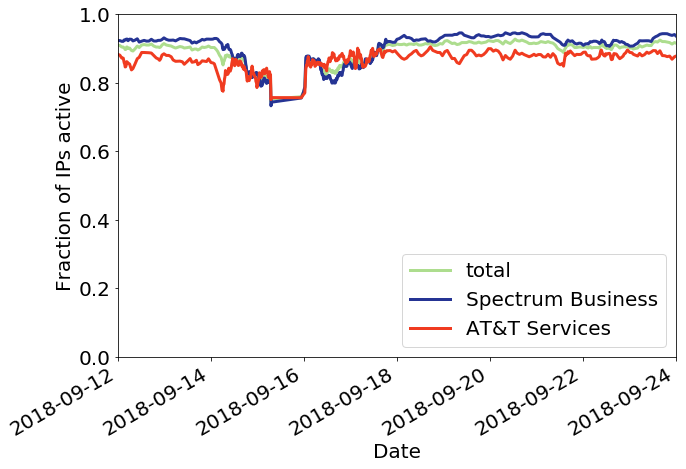}
 \caption{The active fractions for different ISPs in Lancaster, South Carolina}
 \end{subfigure}
 \caption{Two examples of outage detection using DDII versus residential IP scanning, as well as utility records during Hurricane Florence. (a,d) show IP response rate, outage threshold for DDII, and utility reports (b,e) is the scanning frequency per county which is automatically adapted by DDII during failure detection. Finally, (c,f) represents the active fraction of IP addresses within different ISPs. }
 \label{fig:outage-examples}
\end{figure*}

\subsection{Power Outage Detection Accuracy}

As Figure~\ref{fig:avg_score_all} demonstrated, ICMP pings have an expected fraction of success that lies below the fraction of customers receiving power. Hence, to perform an accurate comparison between DDII's predictions of the outages and those as reported by the power company, a normalization needs to take place. When declaring an outage of a certain size (e.g., 20\% of customers losing power), DDII's threshold computations need to be appropriately normalized.

\subsubsection{Detection thresholds}
 We define an outage threshold for utility company data to be the percentage of customers losing power. Specifically, a threshold of $x \%$ in a county means at least $x \%$ of utility customers have lost power at that time. The utility data is updated every 10 minutes so if reporting is correct, it should be observable within 10 minutes of the outage occurring. 





\begin{figure}[t!]
 \centering
 \begin{subfigure}[t]{\linewidth}
 \centering
 \includegraphics[width=\linewidth]{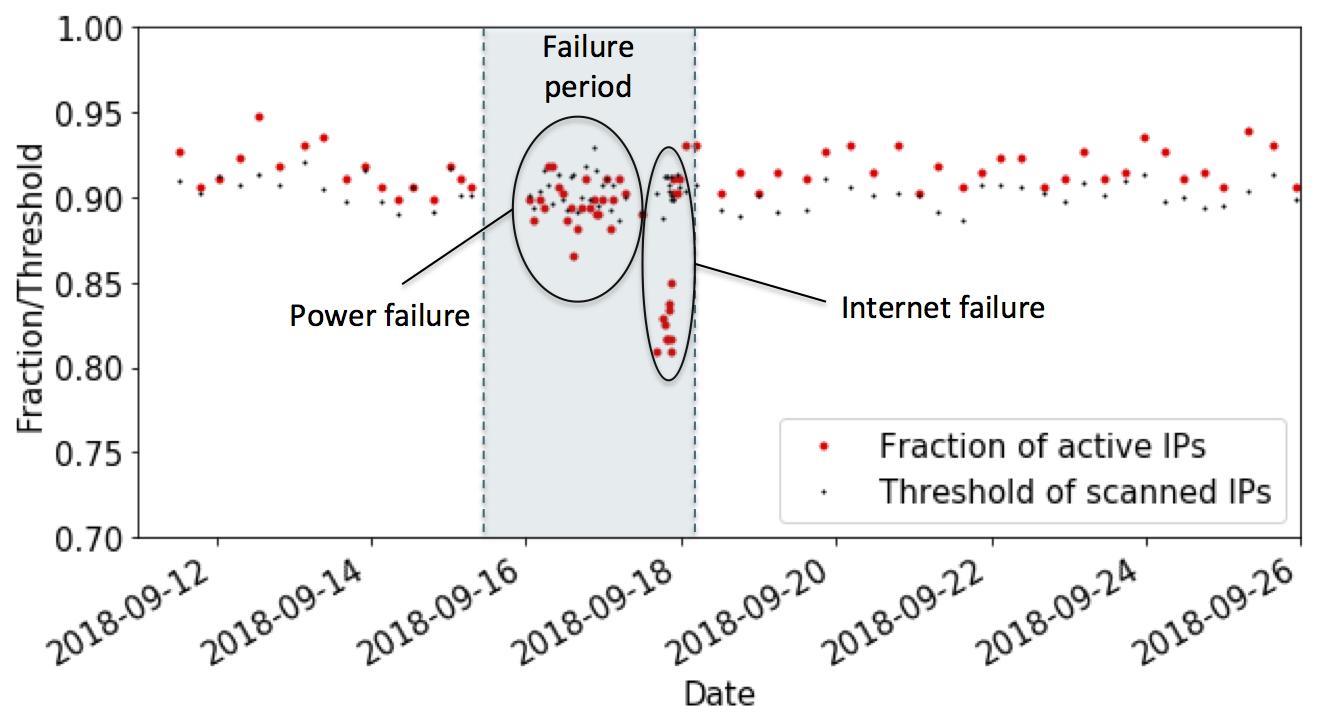}
 \caption{Scanning result for the entire watchlist}
 \end{subfigure} 
 \begin{subfigure}[t]{\linewidth}
 \centering
	\includegraphics[width=\linewidth]{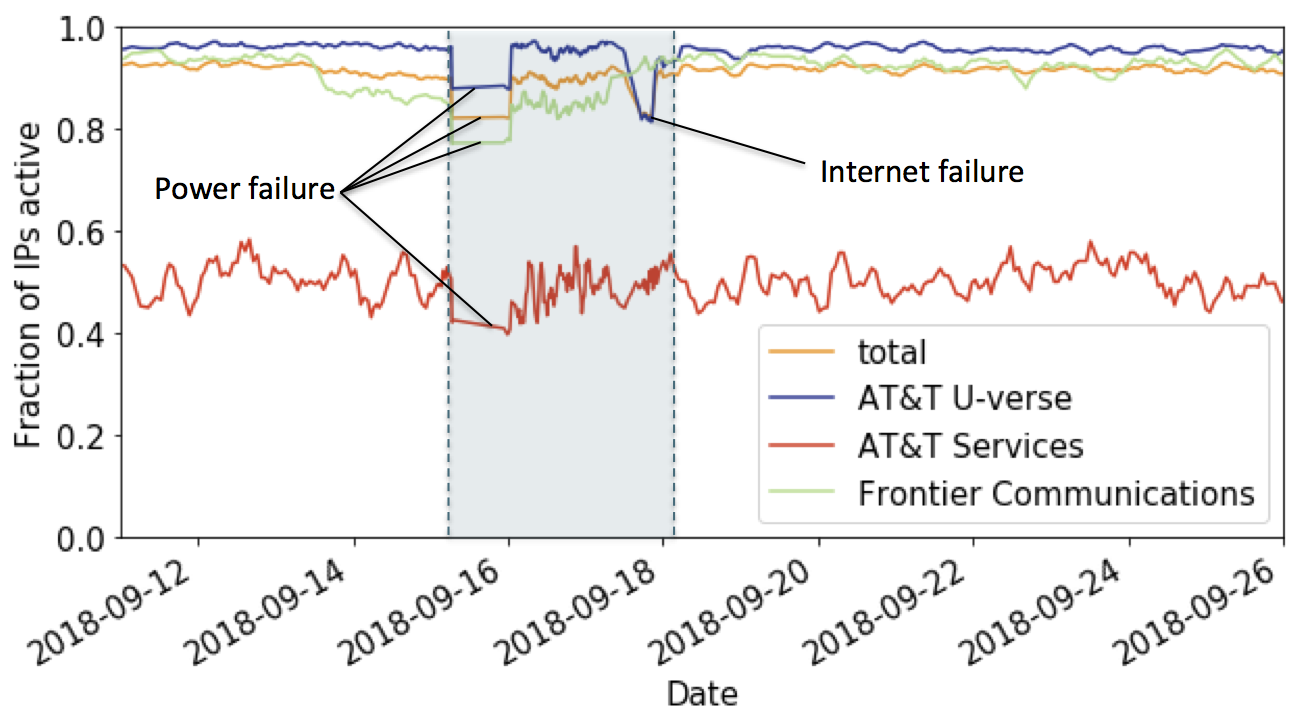}
 \caption{Scanning result per ISP}
 \end{subfigure}
 \caption{Power failure vs Internet failure in Haywood, North Carolina, where the reliable watchlist and the residential watchlist have sizes of $264$ and $6082$, respectively. The scanning frequency increases upon each detection of failure (from roughly every $10$ minutes up to a maximum of every $2$ minutes). This can be observed by denser data points during the failure period.}
 \label{fig:power_Internet}
\end{figure}

Figure \ref{fig:outage-examples} depicts two examples of how DDII metrics compare to the outage statistics reported in utility records for Robeson and Lancaster counties as a function of time over 12 days. Using DDII, as the average IP response rate begins to drop in those counties, we detect that the areas are suspicious of an outage (\ref{fig:outage-examples} (a,d)). We can see that both of these counties can detect true outage using the reliable watchlist. 
In Robeson, the utility data measured that the maximum fraction of affected customers in the monitoring period were $~95\%$ of $27982$ total customers. In Lancaster, when the utility data has its maximum drop, $~50\%$ of $4913$ total customers are affected. Despite different magnitudes of outages, the DDII method was able to detect an outage in both. The number of residential IP addresses in Robeson and Lancaster is $19405$ and $1708$, and the value of $\mathcal{E}$ for them initially is $138$ and $370$, respectively. We can see in this example how sampling a subset of the full residential address space is sufficient for detecting outages. 

For the same counties, Figures \ref{fig:outage-examples}(b,e) illustrate the scanning frequency of IP addresses using DDII. We observe that when DDII detects a failure, the scanning frequency will is increased and sustains a higher frequency with the maximum rate of $0.0083s^{-1}$ until the outage passes. Since in Robeson, the outage lingers for longer, the scanning frequency is high for longer. Furthermore, since the result of our IP probing in the recovery phase is not constantly above DDII threshold, we observe some variation in the frequency during the recovery until it becomes stable. 

Finally, Figures \ref{fig:outage-examples}(c,f) illustrate the difference in outages split by Internet Service Providers (ISP) for the ISPs in each of these counties. We can conclude that because multiple ISPs experienced an outage at the same time that the outage was not due to an Internet outage in one of the ISPs. This allows us to conclude that these were true outages and not issues with a particular ISP.

Figure \ref{fig:power_Internet}(a) depicts the case where our system detects a failure and consequently increases the scanning frequency. A failure starts when at each probing session the fraction of active IPs falls below a unique threshold defined for the scanned set of IPs, which can be different in each probing session. In Figure \ref{fig:power_Internet}(b), we observe both a power failure and an Internet failure in the failure period, in which the former impacts all ISPs, while the latter is only caused by AT\&T U-verse. The power failure event is confirmed from the utility data on Sep 16, 2018, at which point a peak of $7\%$ of tracked customers lose power, while there where no considerable reported outage on Sep 17 to Sep 18. Note that the y-scale is different in two figures for better visualization.

\subsection{DDII Validity}

\begin{figure}[t!]
 \centering
 \begin{subfigure}[t]{\linewidth}
 \centering
 \includegraphics[width=0.95\linewidth]{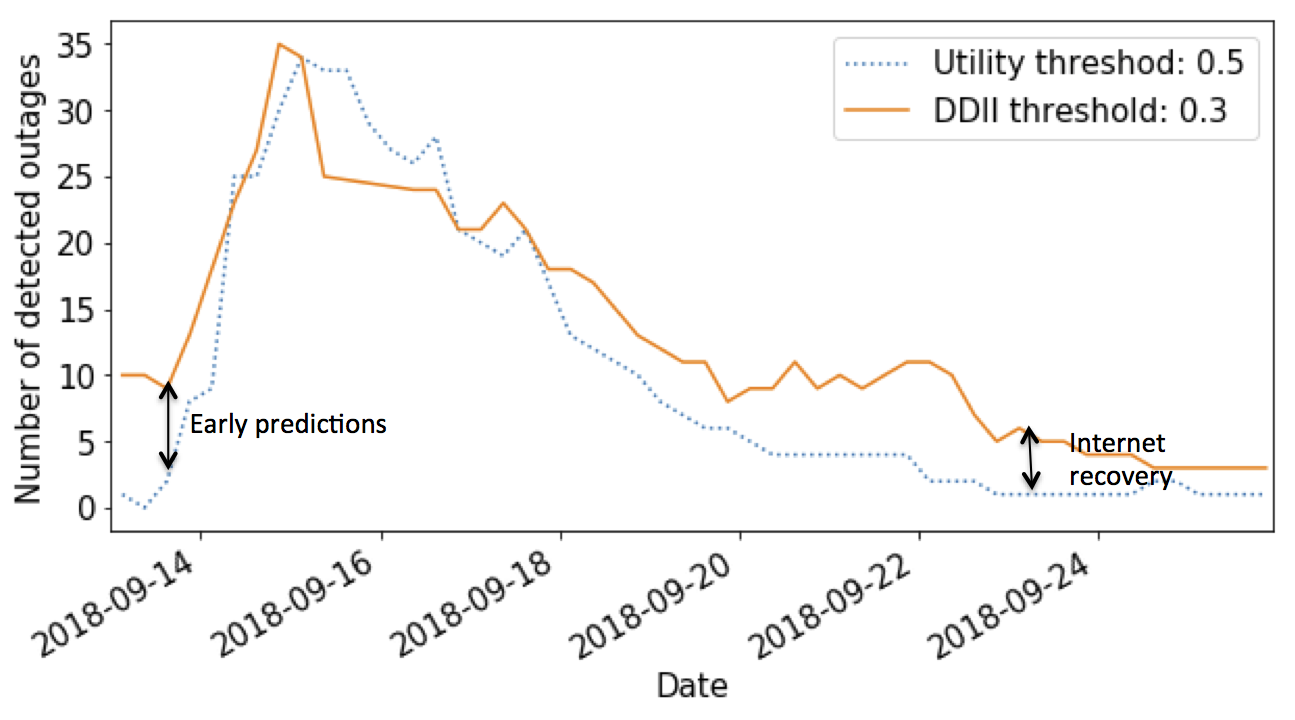}
 \caption{At least 50\% of utility customers reported outage. }
 \end{subfigure} 
 \begin{subfigure}[t]{\linewidth}
 \centering
	\includegraphics[width=0.95\linewidth]{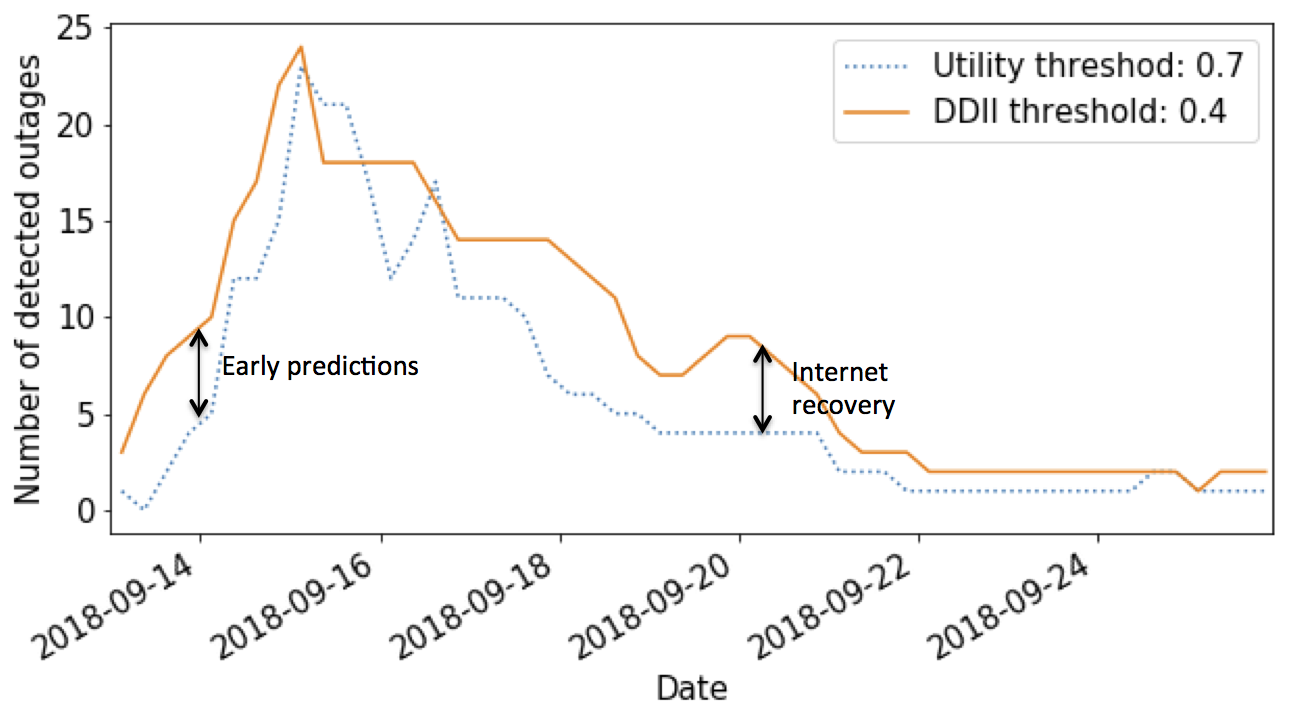}
 \caption{At least 70\% of utility customers reported outage.}
 \end{subfigure}
 \caption{The number of counties with an outage reported by utility data versus detected by our method with different thresholds. These numbers are measured during Hurricane Florence in North Carolina and South Carolina. }
 \label{fig:outage_thresholds}
\end{figure}

To begin, let's consider the case of a major outage to compare the number of detected outages in the utility data and using the DDII method. Figure \ref{fig:outage_thresholds} depicts the number of outages reported by utility companies and detected by our method during Hurricane Florence in North and South Carolina. The legends of $0.5$ and $0.7$ in utility reports mean at least $50\%$ and $70\%$ of customers have reported a power loss, respectively. Whereas the numbers $0.3$ and $0.4$ corresponding to the DDII method mean that the fraction of active IP addresses falls $0.3$ and $0.4$ below the IP monitoring threshold, which as mentioned earlier is defined as: $$\text{average\ score\ of\ scanned\ IP\ addresses }- 0.07$$ We observe that the number of detections in our method is higher than the number of outages measured from utility reports in the beginning and end of the power failure. The former could imply the evacuation of the area and less Internet activity. This could add value to our method by sensing an outage in its early stage, while the latter could imply slower recovery of the Internet connections rather than utility reports, which could be caused by people who have left the area during the storm and have turned their Internet connections off. The widened window of our method is one of the main advantages as we are able to detect an outage before other methods are able to identify the outage. 

\begin{figure}[t!]
 \centering
 \begin{subfigure}[t]{\linewidth}
 \centering
 \includegraphics[width=\linewidth]{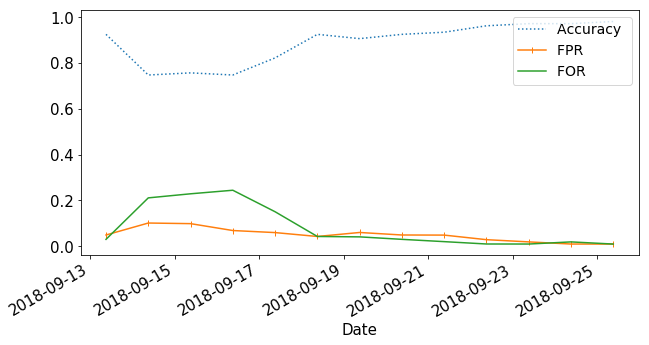}
 \caption{Utility threshold = $0.5$ and DDII threshold = $0.3$. }
 \end{subfigure} 
 \begin{subfigure}[t]{\linewidth}
 \centering
	\includegraphics[width=\linewidth]{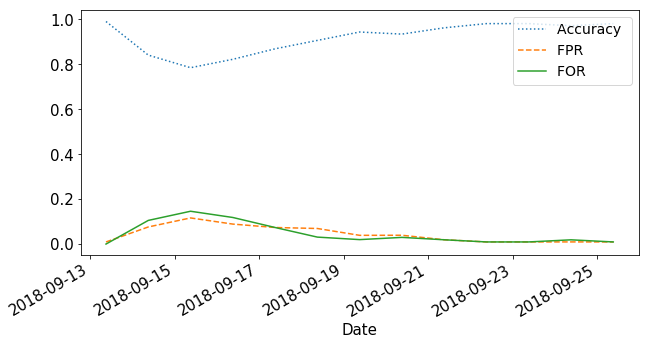}
 \caption{Utility threshold = $0.7$ and DDII threshold = $0.4$. }
 \end{subfigure}
 \caption{The confusion matrix measures during Hurricane Florence in North Carolina and South Carolina. The buffer period is set to $6$ hours. }
 \label{fig:confusion_measure}I
\end{figure}

\begin{figure}[t]
 \centering
 \centering
 \includegraphics[width = \linewidth]{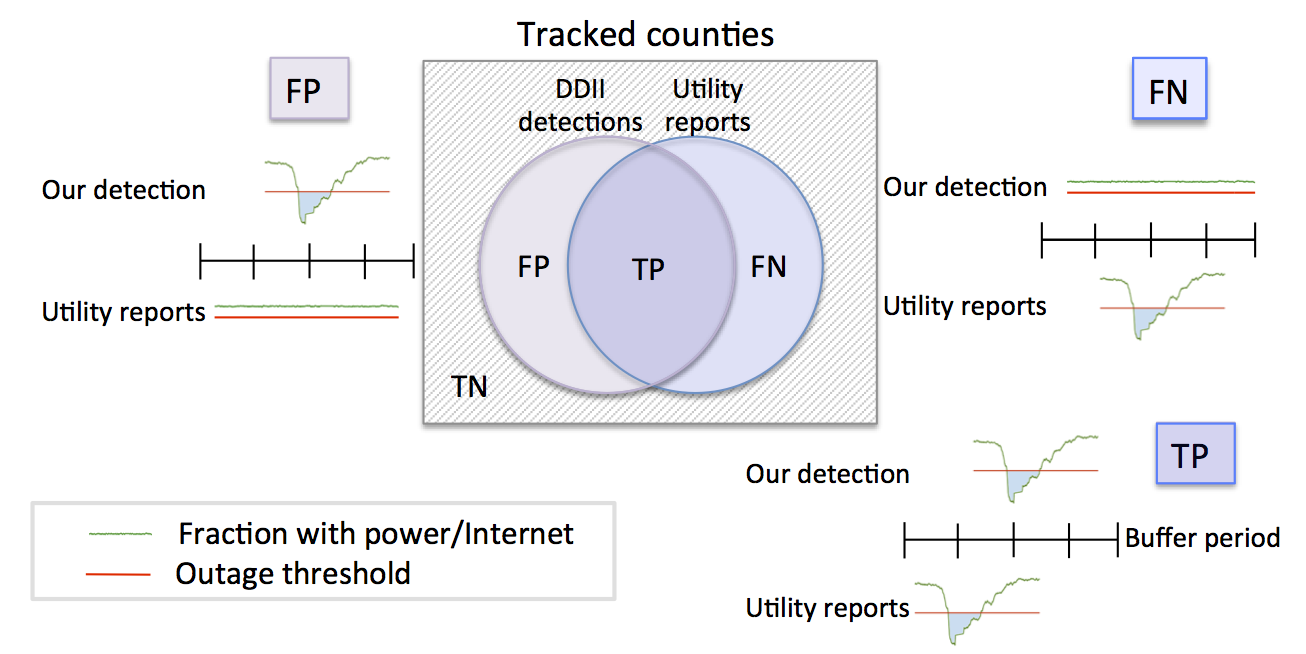}
 \caption{The Venn diagram of the confusion matrix measures and how they are defined in our evaluation. A true detection is defined as when our outage detection and the utility outage report are less than a buffer period apart. }
 \label{fig:Venn}
\end{figure}

To measure the reliability, we compute four parameters of the confusion matrix---a table that is often used to describe the performance of a classification model. We calculate False Positives (FP), False Negatives (FN), True Positives (TP), and True Negatives (TN). Figure \ref{fig:Venn} illustrates how these measures are defined in our evaluation. FP denotes the number of counties where we report a false alarm, whereas FN is the number of counties where we miss an outage event reported by utility companies. On the other hand, TP denotes the number of overlaps between our detection and the outages in the utility reports. Finally, TN includes the rest of the tracked counties in the regions of study.

Using these parameters, we compute accuracy, false positive rate (FPR), and false omission rate (FOR). The definitions are summarized in Table ~\ref{Tab:confusion}.

Note that lower FPR indicates a smaller amount of false alarms compared to $N$, and lower FOR indicates a smaller amount of missed outages compared to non-outage events. Although smaller values for these measure indicates better detection, the larger accuracy indicates a higher fraction of true detections using the DDII method.

\begin{table}[t]
\centering
\bgroup
\def\arraystretch{1.5}
\begin{tabular}{c| c} 
metric & definition \\
\hline
$accuracy$ & $\frac{TP + TN}{P + N} = \frac{TP + TN}{TP + TN + FP + FN}$ \\
\hline 
$FOR$ & $ \frac{FN}{FN + TN} $ \\
\hline
$FPR$ & $ \frac{FP}{N} = \frac{FP}{FP+TN} $ \\
\end{tabular}
\egroup
\caption{Metrics used in our evaluation}
\label{Tab:confusion}
 \end{table}

Figure \ref{fig:confusion_measure} illustrates the measures in Table \ref{Tab:confusion} during Hurricane Florence in North and South Carolina, with a buffer period of $6$ hours. This buffer period means that TP is included as long as the outage detected using our method and the outage detected using the utility data occur within $6$ hours. From Figure \ref{fig:confusion_measure}, we notice that during the peak of the outage, FOR and FPR are slightly higher and accuracy is slightly lower. This is mainly because when the number of outages increases, TN decreases, and with more outages, it is possible to have some of them mislabelled. Once the outages decrease, our metrics return to more stable levels. Also, note that some of the false positives and false negatives could be due to some shortcomings from the utility reports \cite{utilitylimitations}.

\begin{figure}[t!]
 \centering
 \begin{subfigure}[t]{\linewidth}
 \centering
 \includegraphics[width=\linewidth]{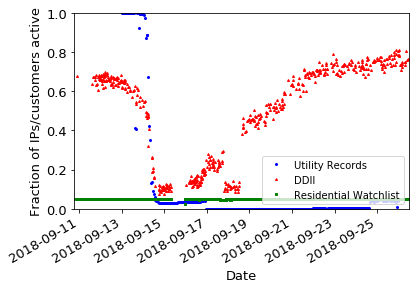}
 \caption{Columbus, North Carolina }
 \end{subfigure} 
 \begin{subfigure}[t]{\linewidth}
 \centering
	\includegraphics[width=\linewidth]{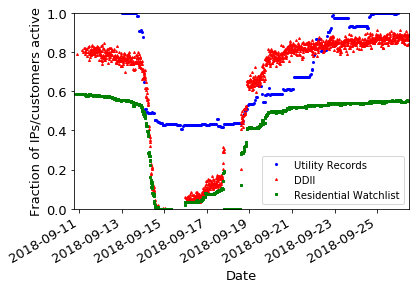}
 \caption{Jones, North Carolina}
 \end{subfigure}
 \caption{The power outage caused by Hurricane Florence in Columbus and Jones counties. The IP probing through DDII visually seems to track power status more precisely than utility reports and residential watchlist in these two cases. }
 \label{fig:Columbus}
\end{figure}

For instance, Columbus is one of the counties for which we detected a false negative during the recovery period, where the number of tracked utility customers, the residential watchlist size, and the reliable watchlist size are $23557$, $5566$, and $313$, respectively. Figure \ref{fig:Columbus}(a) illustrates how the recovery occurs both in DDII and in utility reports. We observe that the utility data \cite{poweroutages} we obtained does not track the recovery phase and continues to report an outage, whereas our method shows the power is going back on. Therefore, we constantly have a false negative in that county after the recovery. This also highlights an advantage of our method---if the reliable IPs are responding to probes at such a high rate, then we can safely determine the power outage has ended, yet the utility records inaccurately shows an outage. Again, we can see the advantage of our method in more precisely measuring outage windows. 

On the other hand, Figure \ref{fig:Columbus}(b) illustrates one of the examples where we had a false positive in our method as the fraction of customers with power does not fall below $70\%$ during the hurricane in Jones county. The number of tracked utility customers, the residential watchlist size, and the reliable watchlist size in Jones are $7095$, $3328$, and $327$, respectively. However, DDII results indicate that more customers have lost power than what is derived from utility reports. Given the knowledge of Hurricane Florence and the fact that both the residential and reliable watchlists reach a response rate close to $0\%$, we can most likely conclude that a significant outage occurred. While there is no easy way to answer which of the results are more accurate, we believe our method has the ability to add to what power utility reports have to present.

\section{Related Work}\label{RelatedWork}

The works most closely related to ours are \cite{schulman2011pingin, quan2013trinocular, richter2018advancing}. Schulman and Spring in \cite{schulman2011pingin} use ICMP pings to determine the responsiveness of IP addresses during thunderstorms. Called Thunderping, it checks each IP from 10 different vantage points to asses if the IP address is down. 
Padmanabhan et al. \cite{padmanabhan2019find} further use Thunderping data to detect Internet failure events that affect multiple users simultaneously, and show that dependent disruption events do not always affect entire /24 address blocks and can therefore be missed by prior techniques
Furthermore, Aceto et al. provide a comprehensive survey on Internet outage detection methods\cite{acetosurvey}. Even though there is some existing research analyzing weather-caused Internet failures, which could be potentially generalized to power failure detection, they have some shortcomings we try to address. The difference between previous work and ours is first, we differentiate between power and Internet failure. Second, while they monitor all IP addresses known as residential, we dynamically determine how informative an IP address is on power status as an individual, regardless of their /24 blocks. Finally, they only monitor IP addresses upon knowledge of a thunderstorm to observe their behavior, while we build a system to actually detect the power failure in their early stage and focus in on areas with suspicious behavior by increasing the probing frequency in those areas. 

Trinocular in \cite{quan2013trinocular} introduces an Internet monitoring system that aims to consistently detect Internet outages in small, ``edge" networks. They do so through active probing, where they use probes driven by Bayesian inference to learn the current status of the Internet. However, false positives in a few address blocks can dominate and Trinocular's outage detection must be filtered for most events to be correct. Unlike Trinocular, Richter et al. \cite{richter2018advancing} and Dainotti et al. \cite{Dainotti2012} introduce a passive probing to detect Internet failures. Their approach focuses on offline detection of disruptions in CDN log files and analyzing Internet Background Radiation (IBR) traffic respectively. Furthermore, Shah et al. \cite{shah2017disco} use existing long-running TCP connections to identify bursts of
disconnections and use power outage in Amsterdam as one of their study cases in a small scope.
The main difference between our work and their methods is that they do not cover online detection
, while our method aims to detect failures in real-time. Moreover, these works may not be sensitive enough to detect small outage events, and finally, their main focus is on Internet outages rather than the impact of power outages on the Internet.


C. k\"{o}pp in \cite{kopp2013controlled} introduces an analysis of using Border Gateway routing to measure Internet outages. Their methodology analyzes BGP data dumps and matches outages to IP prefix geolocation. They show how their methodology performs through case studies of past power outages, specifically in Egypt and New Zealand. In addition, \cite{Dainotti2014} detected country-wide internet outages caused by government censorship by analyzing BGP control and data plane traffic. However, since BGP is a routing protocol, they were only able to detect large scale Internet outages. This is because Internet Service Providers put in a lot of effort to make sure that their routing infrastructures do not go down, even when residential and local business Internet access goes down. In the case of a power outage, even if Internet usage does go down, BGP routing might still be online.

 Cardona et al. attempt to find if short term weather patterns, like snow or rain, affect Internet usage and traffic demand \cite{cardona2013weather}. They found that the impact of precipitation was not uniform. While they were unable to prove that there was a dependency between weather and Internet traffic, they believe that there is some correlation. However, they did not address the effect of power outages on the activity of IP addresses.

Casillo et al. provide a comprehensive survey in power systems research \cite{castillosurvey}. According to the survey, the majority of the work related to power systems has been done in developing statistical models to forecast power outages caused by natural disasters such as hurricanes, severe storms, and heatwaves. 
\cite{Nateghi2014, stormvariables}. Another major area of focus in power outage research is simulating and understanding the impact of contingency-based outages, which usually result from cascading failures \cite{castillosurvey}. These statistical models and simulations give a good idea in advance of the severity and impact on infrastructure due to power outages and allow the utility companies to perform power restoration planning and identify opportunities to make the power grid more resilient. However, these methods are not Internet-based. As our results show, monitoring the Internet to detect outages is cheap, can provide us with extensive information, and it is not prone to human error or direct attacks to power grid monitoring systems. 

Information from social media combined with measurements in power distribution systems has also been used to detect power outages. Sun et al. use real-time tweets and a probabilistic framework integrating textual, temporal and spatial information to detect the outage \cite{suntweets}. Sevlian et al. use a detection method based on real-time load and line flow measurements in power distribution systems, which requires placing sensors at optimal locations in the distribution system to reduce mean detection error probability \cite{measurementsloadline}. While these methods are very useful, they can only serve as complementary data on power status since they are based on human reports which are prone to human error. 

Wireless sensor networks (WSN) can be used to realize low-cost embedded electric utility monitoring and diagnostic systems \cite{akyildiz2002wireless, bello2007design, garcia2004reconfigurable, gungor2006survey, lu2009online, yang2007survey}. However, WSNs themselves lack the reliability and security of wired network nodes, require constant recharging, and communication speeds are comparatively lower than wired networks \cite{bhattacharyya2010comparative}.

\section{Discussion and Future Work}\label{Discussion}
In this work, we present a design study of a power grid monitoring system based on IP probing. Our approach is principled, using a simple outage-centric model of the Internet that learns the current status of the power from Internet probes. We measure the correlation between power company data and our data-driven from the Internet and adjust our parameters so that the results of our IP probing better match power company data.

\subsection{Minor outage events}
Our evaluation analyzed power outages for major weather events such as hurricanes, which cause widespread impact affecting multiple counties. While DDII can be applied to identify power outages for minor events, it requires further fine-tuning to reduce the false positive and false negative rates. Specifically, based on \ref{sec:reliableIPsizing}, further evaluations of minor events are required to answer questions like: What is the minimum number of samples within an iteration for a region for accurate detection? Does this number vary across counties?

\subsection{Real-time outage map}
As part of ongoing work, we will implement the design that we studied in this paper as a real-time live map that shows our estimate of the power system status nationwide. This could potentially allow power companies to investigate counties that may experience outages in their early stages.

\subsection{Learning IP behavior}
Another avenue we are exploring includes studying behavior of IP addresses among different counties more extensively to map a specific behavior of IP addresses within the reliable watchlist in each county to a power outage percentage. As a part of this direction, we can apply machine learning methods to the reliable watchlist for this purpose, using features such as average score of IPs scanned, average score of IPs within the county, county population, number of IPs monitored in the county, distance from average score and fraction of active IPs, and similar information extracted from different ISPs.



\bibliographystyle{abbrv}
\bibliography{main}

\end{document}